\documentclass[english,nocopyrightspace]{sigplanconf}
\usepackage[T1]{fontenc}
\usepackage[latin9]{inputenc}
\usepackage{verbatim}
\usepackage{amsthm}
\usepackage{amsmath}
\usepackage{amssymb}
\usepackage{graphicx}

\makeatletter
\theoremstyle{plain}
\newtheorem{thm}{\protect\theoremname}
\theoremstyle{definition}
\newtheorem{example}[thm]{\protect\examplename}
\newcommand{\code}[1]{\texttt{#1}}

\@ifundefined{showcaptionsetup}{}{%
 \PassOptionsToPackage{caption=false}{subfig}}
\usepackage{subfig}
\makeatother

\usepackage{babel}
\usepackage{listings}
\providecommand{\examplename}{Example}
\providecommand{\theoremname}{Theorem}

\begin{document}

\title{Towards Taming Java Wildcards and Extending Java with Interval Types}

\authorinfo{Moez A. AbdelGawad}
{Informatics Research Institute, SRTA-City, Alexandria, Egypt}
{\texttt{moez@cs.rice.edu}}
\maketitle
\begin{abstract}
\global\long\def\pcgp{\ltimes}
Of the complex features of generic nominally-typed OO type systems,
wildcard types and variance annotations are probably the hardest to
fully grasp. As demonstrated when adding closures (\emph{a.k.a.},
lambdas) and when extending type inference in Java, wildcard types
and variance annotations make the development and progress of OO programming
languages, and of their type systems in particular, a challenging
and delicate task.

In this work we build on our concurrent work, in which we model Java
subtyping using a partial graph product, to suggest how wildcard types
in Java can be generalized, and simplified, to \emph{interval types}.
In particular, interval types correspond to endpoints of paths in
the Java subtyping graph.

In addition to being a simple and more familiar notion, that is easier
to grasp than wildcard types, interval types are strictly more expressive
than wildcard types. As such, we believe interval types, when developed
and analyzed in full, will be a welcome addition to Java and other
similar generic nominally-typed OO programming languages.
\end{abstract}


\section{\label{sec:Introduction}Introduction}

Java wildcard types~\cite{JLS05,JLS14,Torgersen2004,MadsTorgersen2005}
(also known as `Java wildcards') are wild. Wildcard types in Java
are an instance of so-called \emph{usage-site} variance annotations.
Another form of variance annotations, called \emph{declaration-site
}variance annotations, is supported in other generic nominally-typed
OO programming languages such as C\#~\cite{CSharp2015}, Scala~\cite{Odersky14}
and Kotlin~\cite{Kotlin18}. Declaration-site variance annotations
are (almost) as wild as usage-site variance annotations.

\subsection{Motivating Example}

To illustrate how wild Java wildcard types, and variance annotations
more generally, can be, consider the following Java code. The reader
is invited to tell (before he or she checks the answer below) which
of the variable assignments in the following code---all involving
wildcard types---will typecheck and which ones will the Java typechecker
complain about.

\begin{lstlisting}[language=Java,basicstyle={\small\ttfamily},frame=lines]
class E<T> {}
class C<T> extends E<E<T>> {}
class D<T> extends C<D<T>> {}

class App {
  public static void main(String[] args){
    C<? extends D<?>> cxd = null;
    C<D<?>> cd = null;
    D<?> d = null;
    D<D<?>> dd = null;
    D<? extends D<?>> dxd = null;

    // Assuming the decls above typecheck,
    // which of the following eleven assign-
    // ments typecheck and which ones don't?
    // ... and why?? (The answer involves
    // figuring out the correct subtyping
    // relations between the types of the
    // variables in the assignments).
    //

    cxd = cd;      cd = d;        cxd = d;

    d = dd;        dd = d;

    dxd = dd;      dd = dxd;

    d = dxd;       dxd = d;

    cxd = dxd;     cd = dxd;
  }
}
\end{lstlisting}

Answer: All class and variable declarations in the code above typecheck,
but the second assignment in each of the five groups of variable assignments
(\emph{i.e.}, elements of the middle column) does not typecheck, while
the other six assignments will typecheck.

Now, did the reader figure this answer correctly? ... and quickly?
... and, most importantly, can the reader tell the \emph{why} for
his or her findings? ... (Unfortunately, the error messages that the
Java compiler emits, which involve wildcard ``capturing'', are largely
unhelpful in this regard).

The difficulty in figuring out which of the assignments above (let
alone assignments involving even more complex generic types) do typecheck
and which ones do not, as well as the difficulty in figuring out the
reason in each case, and the almost total unhelpfulness of error messages,
is an example of how hard (\emph{a.k.a.}, wild) reasoning about wildcard
types can be.

A main motivation for our research, and for the work presented in
this paper in particular, is to face this situation head-on and try
to significantly improve it. We believe the research presented here
and in other closely-related publications is a significant step towards
that goal. (Based on earlier research we did~\cite{AbdelGawad2016c,AbdelGawad2017b}
that largely complements the work we present in this paper, we believe
Java generics error diagnostics, for example, can be significantly
improved once wildcard types are fully tamed and generalized to interval
types.)

This paper is structured as follows. In Section~\ref{sec:Background}
we briefly discuss the background needed for reading this paper. In
Section~\ref{sec:Type-Intervals-Constructing} we then formally define
OO interval types and the construction method of the Java subtyping
relation with interval types. In Section~\ref{sec:Examples} we present
examples that illustrate the definitions in Section~\ref{sec:Type-Intervals-Constructing}
(In Appendix~\ref{sec:SageMath-Code} we present the SageMath code
we developed to help us in generating these examples). In Section~\ref{sec:Related-Work}
we briefly discuss research related to the research we present in
this paper. In Section~\ref{sec:Discussion} we make some concluding
remarks and discuss some possible future work that can build on work
presented in this paper.

\section{\label{sec:Background}Background}

The background necessary for reading this paper is basically the same
as that of~\cite{AbdelGawad2017a} (A summary is presented in Section~2
of~\cite{AbdelGawad2018b}). In the following sections of this paper
we largely adopt the same assumptions, vocabulary and notation of~\cite{AbdelGawad2017a,AbdelGawad2018b}.

In particular, the construction method we use to construct the Java
subtyping relation with interval types (which is formalized in Section~\ref{sec:Type-Intervals-Constructing}
of this paper), except for its use of interval types, is the same
as the construction method used to construct the Java subtyping relation
with wildcard types (as presented in~\cite{AbdelGawad2018b}).

\section{\label{sec:Type-Intervals-Constructing}Interval Types and Constructing
The Java Subtyping Relation}

In this section we formally introduce interval types, then we formally
define the construction method of the Java subtyping relation between
generic reference types with interval type arguments.

\subsection{Object-Oriented Interval Types}

Informally, similar to closed intervals over real numbers or over
integers, which are sets of numbers between and including two bounding
numbers, intervals over a directed graph are sets of vertices between
and including two bounding vertices in the graph.\footnote{The two bounds of a graph interval are usually called its \emph{lowerbound}
and its \emph{upperbound}, particularly when the graph is that of
a partially-ordered set, as is the case for the OO subtyping relation
and for the smaller-than-or-equals relation on numbers.}

Formally, we define intervals of a directed graph as the quotient
set of the set of paths of the graph (including trivial, zero-length
paths from vertices of the graph to themselves) over the equivalence
relation of paths of the graph with the same endpoints, as follows.

Let $G=(V,E)$ be a directed graph, and let $P(G)$ be the set of
paths in $G$. For a path $p$ in $P(G)$ let $s(p)\in V$ and $e(p)\in V$
denote the (possibly equal) start and end vertices of $p$ (\emph{i.e.},
the endpoints of path $p$). For paths $p_{1}$ and $p_{2}$ in $G$
(\emph{i.e.}, $p_{1},p_{2}\in P(G)$), define the relation $\leftrightarrow$
over $P(G)$ as the equivalence relation
\[
p_{1}\leftrightarrow p_{2}\mathrm{\quad iff\quad}s(p_{1})=s(p_{2})\wedge e(p_{1})=e(p_{2}).
\]
In agreement with intuition, using the properties of $=$ and conjunction
($\wedge$) it is easy to show that relation $\leftrightarrow$ is
a reflexive ($\forall p\in P(G),p\leftrightarrow p$), symmetric ($\forall p_{1},p_{2}\in P(G),p_{1}\leftrightarrow p_{2}\implies p_{2}\leftrightarrow p_{1}$)
and transitive relation ($\forall p_{1},p_{2},p_{3}\in P(G),p_{1}\leftrightarrow p_{2}\wedge p_{2}\leftrightarrow p_{3}\implies p_{1}\leftrightarrow p_{3}$),
\emph{i.e.}, that $\leftrightarrow$ is an equivalence relation. Relation
$\leftrightarrow$ thus partitions $P(G)$.

The set $I(G)$ of intervals of graph $G$ is then defined as the
quotient set
\[
I(G)=P(G)/\leftrightarrow
\]
(\emph{i.e.}, as the set of equivalence classes defined by $\leftrightarrow$).

Equivalently, intervals of $G$ can be defined using the \emph{reflexive
transitive closure} of $G$. That is because the intervals of $G$
are in one-to-one correspondence with the edges (including self-edges/loops)
of $RTC(G)$, the reflexive transitive closure of $G$. In other words,
if $RTC(G)=(V,E_{rtc})$ then we have 
\[
I(G)\simeq E_{rtc}.
\]

Hence, based on the definition of graph intervals, one graph interval
can correspond to multiple (one or more) paths of the graph, all having
the same endpoints, but one path of the graph corresponds to exactly
one graph interval, namely the interval defined by the endpoints of
the path (\emph{i.e.}, its start and end vertices~\cite{Hammack2011}).\footnote{We call paths corresponding to a graph interval its `witnesses'.
More precisely, similar to formal proofs of valid logical statements
in formal logic, we call a path in a graph a \emph{witness} to the
graph interval defined by the endpoints of the path. In fact we found
some resemblance between graph intervals in graph theory (as we define
them in this paper) and valid logical statements in formal mathematical
logic. In formal logic a logical statement must have a formal proof
for the statement to be a valid logic statement. Similarly, a pair
of graph vertices must have a path connecting its two vertices for
the pair to define a graph interval. Further, in formal logic a valid
logical statement can have multiple (one or more) witnessing proofs.
Likewise, a graph interval can have multiple (one or more) witnessing
paths. Also, the $\leftrightarrow$ relation between graph paths is
analogous to the equivalence of proofs with the same premises and
conclusions (`proof irrelevance'), but, other than noting the analogy
in this footnote, we do not explore or take the analogy any further
in this paper.} As such, using two standard notations for intervals, we denote a
graph interval $I$ with endpoints $v_{1}$ and $v_{2}$ either by
$I=[v_{1}-v_{2}]$, or sometimes equivalently by $I=[v_{1},v_{2}]$
(given that we define intervals over \emph{directed} graphs, the order
of the vertices in a graph interval expression matters).

Informally, an interval $I=[v_{1}-v_{2}]$ over a graph $G$ can be
viewed as a pair of vertices $(v_{1},v_{2})$ in $G$ where a path
from $v_{1}$ to $v_{2}$ is \emph{guaranteed} to exist in $G$ ($v_{2}$
is said to be \emph{reachable} from $v_{1}$). Not all pairs of vertices
of $G$ satisfy this condition. A pair of vertices that do satisfy
this connectedness condition corresponds to an interval over $G$,
while the vertices of a pair that does not satisfy the condition are
called \emph{disconnected} \emph{vertices} (they are sometimes also
called \emph{parallel} \emph{vertices}, particularly when $G$ is
the graph of a partial order and when also the inverse pair $(v_{2},v_{1})$
does \emph{not} form an interval, and are usually denoted by $v_{1}||v_{2}$
in order theory literature). In other words, a graph interval over
a directed graph $G$ corresponds to a pair of \emph{connected} or
\emph{reachable }vertices in $G$, where (in the context of directed
graphs) `connectivity' is understood to be \emph{to} the second mentioned
vertex ($v_{2}$) and `reachability' is understood to be \emph{from}
the first mentioned vertex ($v_{1}$), \emph{i.e.}, we say $v_{2}$
is reachable from $v_{1}$ or, equivalently, that $v_{2}$ is connected
to $v_{1}$.

We then simply define \emph{OO interval types} in an OO type system
as graph intervals over the graph of the OO subtyping relation (\emph{i.e.},
the graph whose vertices are OO types and whose edges correspond to
the subtyping relation between OO types).

\subsubsection{\label{sub:The-Containment-Relation}The Containment and Precedence
Relations}

Similar to intervals of real numbers or intervals of integers, graph
intervals and interval types can be (partially) ordered by a containment
relation, where a graph interval $I_{1}$ is said to be \emph{contained-in}
another interval $I_{2}$ if some path $P_{1}$ corresponding to $I_{1}$
is, in its entirety, a \emph{subpath}~\cite{Hammack2011} of some
path $P_{2}$ corresponding to $I_{2}$ (\emph{i.e.}, for $I_{1}$
to be contained in $I_{2}$ the path $P_{1}$ must share its vertices
with $P_{2}$, in the same order as the vertices occur in $P_{1}$).
If an interval $I_{1}$ is contained in an interval $I_{2}$, sometimes
we call $I_{1}$ a \emph{subinterval} of $I_{2}$ and, equivalently,
call $I_{2}$ a \emph{superinterval} of $I_{1}$.

The definition of the containment relation between graph intervals
may seem a bit convoluted, but it should be intuitively clear. (Visually
speaking, \emph{i.e.}, when presented with the diagram of a graph,
it is usually immediately obvious whether a graph interval is contained
in another.)

In accordance with our definition of interval types, we define the
containment relation between interval types as containment between
their corresponding graph intervals over the graph of the OO subtyping
relation. It should be intuitively clear also that the containment
relation is a partial ordering between interval types, which itself---\emph{i.e.},
the containment ordering---can be modeled by a directed graph (which
can be presented,\emph{ }for example, using the Hasse diagram of the
partial ordering).

Another relation that can be defined on graph intervals (and interval
types accordingly) is the \emph{precedence} relation, where an interval
$I_{1}=[v_{1},v_{2}]$ is said to \emph{precede} an interval $I_{2}=[u_{1},u_{2}]$
if and only if there exists an interval $I_{3}=[v_{2},u_{1}]$ (\emph{i.e.},
if and only if the pair $(v_{2},u_{1})$ actually defines an interval).
In other words, interval $I_{1}$ precedes interval $I_{2}$ if and
only if the end vertex of $I_{1}$ and the start vertex of $I_{2}$
are connected%
\begin{comment}
 (note the reversal of order of the vertices compared to the order
of the intervals)
\end{comment}
. If $I_{1}$ precedes $I_{2}$ it may be equivalently said that $I_{2}$
\emph{succeeds} or \emph{follows }$I_{1}$. Currently\emph{ }(\emph{i.e.},
in this paper), we do not have a particular use or application of
the precedes relation in constructing and modeling the generic Java
subtyping relation, but we do not preclude the possibility---indeed,
we believe---that the precedes relation may be useful in some future
work that builds on the work we present in this paper.

\subsection{Constructing The Java Subtyping Relation with Interval Types}

As we did for wildcard types (see~\cite{AbdelGawad2018b}), to construct
the Java subtyping relation with interval types we use the partial
Cartesian graph product operator $\pcgp$ from~\cite{AbdelGawad2018a}.

First, however, we define a graph constructor $S^{\Updownarrow}$
(similar to our definition of the graph constructor $S^{\triangle}$
in~\cite{AbdelGawad2018b}) that constructs the graph whose vertices
are all the interval type arguments that can be defined for the graph
of a subtyping relation $S$ and whose edges model the containment
relation between these arguments. In other words, operator $S^{\Updownarrow}$
constructs (the graph of) the containment relation between interval
types over $S$ as described in Section~\ref{sub:The-Containment-Relation}.

With $S^{\Updownarrow}$ in hand, the Java subtyping relation with
interval types can now be defined as the solution $S$ of the recursive
graph equation
\begin{equation}
S=C\pcgp_{C_{g}}S^{\Updownarrow}\label{eq:S}
\end{equation}
where $S$ is the graph of the subtyping relation, $C$ is the graph
of the subclassing relation, and $C_{g}$ is the set of generic classes
(a subset of classes in $C$).

As is standard for recursive equations, Equation~(\ref{eq:S}) can
be solved iteratively (thereby formalizing our construction method)
using the equation
\begin{equation}
S_{i+1}=C\pcgp_{C_{g}}S_{i}^{\Updownarrow}\label{eq:Si}
\end{equation}
where the $S_{i}$ are finite successive better approximations to
the infinite relation $S$, and $S_{0}^{\Updownarrow}$ is an appropriate
initial graph of the containment relation between interval types (which
we take as the graph with one vertex, having the default interval
type argument `\code{?}' as its only vertex, and having no containment
relation edges). Equation~(\ref{eq:Si}) thus formally and concisely
describes the construction method of the generic Java subtyping relation
between ground Java reference types with interval types.

\begin{comment}
TODO (see corresponding part in~\cite{AbdelGawad2018b}): $(v,i)[p>=i0>=max(e,3v-3)]->(gi+ng,<=i1<=),v^{2},\#paths$

explain the equations: what they mean and where they come from

\#Edges: $CartProdEdges=m_{1}n_{2}+n_{1}m_{2}$ $PartCartProd=???,TransRed=???$

, $S_{0}=C\pcgp?$
\end{comment}

As a comparison of their defining equations reveals, the main difference
between the construction of the graph of the Java subtyping relation
with interval types (presented here) and the construction of the graph
of the Java subtyping relation with wildcard types (presented in~\cite{AbdelGawad2018b})
lies in using the operator $S^{\Updownarrow}$ in place of operator
$S^{\triangle}$ to construct the type arguments of generic classes.

In~\cite{AbdelGawad2017a,AbdelGawad2018b} we noted that the generic
Java subtyping relation exhibits self-similarity. The equation for
Java subtyping with interval types demonstrates that extending Java
with interval types, as a generalization of wildcard types, preserves
the self-similarity of the subtyping relation. As is the case for
wildcard types, the self-similarity of the Java subtyping relation
with interval types comes from the fact that the second factor (\emph{i.e.},
$S^{\Updownarrow}$, the graph of the containment relation between
interval type arguments) of the partial product $\pcgp$ defining
$S$ is derived iteratively (in all but the first iteration) from
the first factor of the product (\emph{i.e.}, from\emph{ }$C$, the
subclassing relation). The implications (see~\cite[Section 3]{AbdelGawad2018b})
of the dependency of the definition of the subtyping relation on the
definition of the subclassing/class inheritance relation in nominally-typed
OO programming languages, regarding the value of nominal typing and
its influence and effects on the type system of Java and other similar
nominally-typed OO languages, are thus also preserved.

\section{\label{sec:Examples}Examples of Constructing The Java Subtyping
Relation with Interval Types}

In this section we present examples for how the generic Java subtyping
relation between ground reference types with interval type arguments
can be iteratively constructed. As we do in the examples section of~\cite{AbdelGawad2018b},
we use colored edges in the diagrams below to indicate the self-similarity
of the Java subtyping relation.

Also, so as to have the lower bound always on the left of a type argument
expression and the upper bound always on the right of a type argument
expression (as is the customary mathematical notation for intervals)
in the diagrams we use the notation `\code{T~<:~?}' (to mean `\code{T~extends~?}')
instead of `\code{?~:>~T}' (which means `\code{?~super~T}')
to express type arguments upper-bounded by \code{O} (\emph{i.e.},
\code{Object}).

\begin{figure}
\noindent \begin{centering}
\subfloat[$C_{1}$]{\noindent \protect\centering{}\protect\includegraphics[scale=0.5]{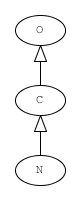}\protect}~~~~~\subfloat[$S_{11}=C_{1}\protect\pcgp_{\{\mathtt{C}\}}S_{10}^{\Updownarrow}$]{\noindent \protect\centering{}~~~~~~~~~\protect\includegraphics[scale=0.5]{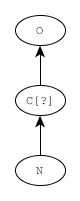}~~~~~~~~~\protect}~~~~~\subfloat[$S_{12}=C_{1}\protect\pcgp_{\{\mathtt{C}\}}S_{11}^{\Updownarrow}$]{\noindent \protect\centering{}\protect\includegraphics[scale=0.3]{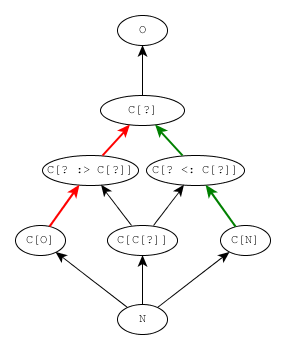}\protect}
\par\end{centering}

\noindent \begin{centering}
\subfloat[\label{fig:IC2}$S_{13}=C_{1}\protect\pcgp_{\{\mathtt{C}\}}S_{12}^{\Updownarrow}$]{\noindent \protect\centering{}\protect\includegraphics[scale=0.3]{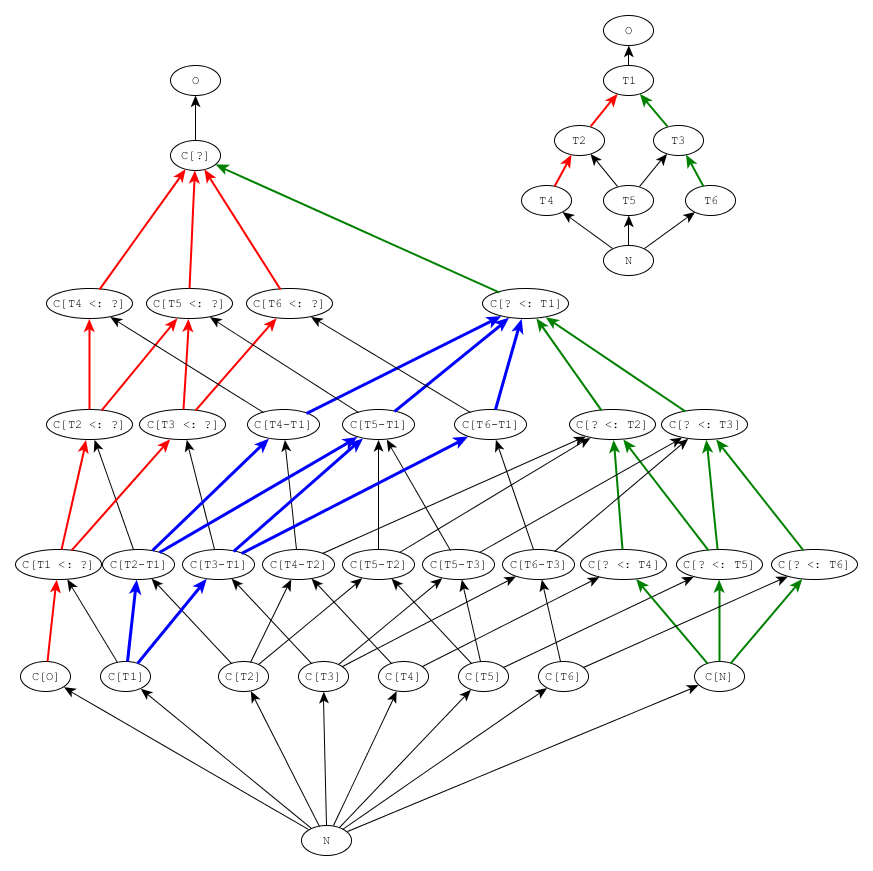}\protect}
\par\end{centering}

\protect\caption{\label{fig:Illustrating-the-use}Constructing generic OO subtyping
with interval types}
\end{figure}

\begin{figure}
\noindent \begin{centering}
\includegraphics[scale=0.3]{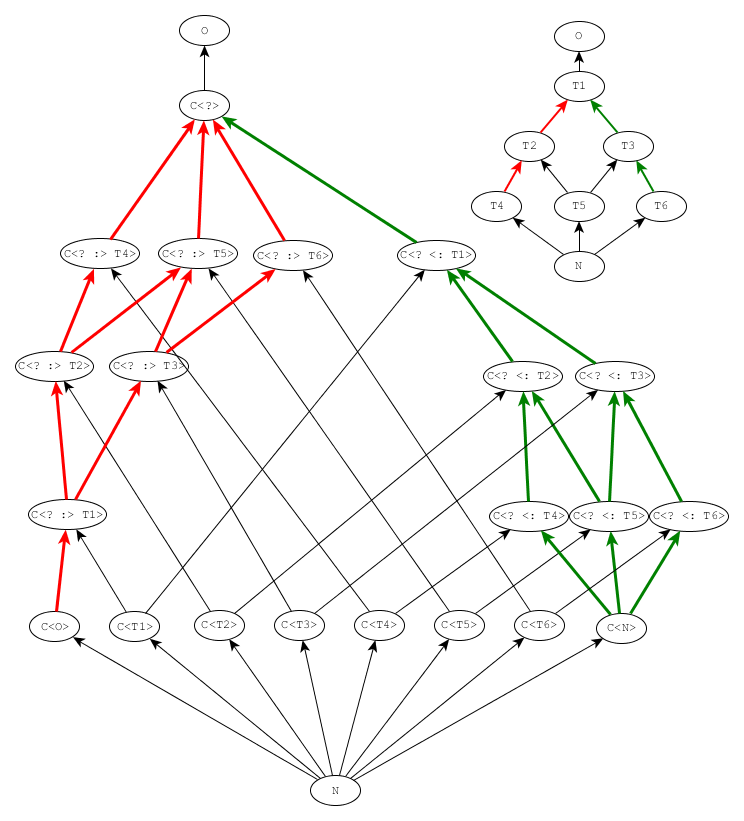}
\par\end{centering}

\protect\caption{\label{fig:Comparing} $S_{13}$ with wildcard types (for comparison)}
\end{figure}

Also, in the diagrams we use wildcards notation (`\code{?}', `\code{?~<:~T}'
and `\code{T~<:~?}') as much as possible, even though unnecessary,
so as to indicate the equivalence of some types that have interval
type arguments to types that have wildcard type arguments. As such,
only interval type arguments that \emph{cannot} be expressed as wildcard
type arguments are expressed using the proper intervals notation (`\code{T1~-~T2}'). 
\begin{example}
Consider the Java class declaration

\code{\textbf{class} C<T>~\{\}}.

\noindent The graphs in Figure~\ref{fig:Illustrating-the-use} illustrate
the construction of the subtyping relation $S_{1}$ corresponding
to this declaration. (In Figure~\ref{fig:IC2}, so as to shorten
names of types in $S_{13}$, we use \code{T1} to \code{T6} to stand
for types of $S_{12}$ other than \code{O} and \code{N}.)

A comparison of Figure~\ref{fig:IC2} to Figure~\ref{fig:Comparing}---which
is a reproduction of Figure~2d of~\cite{AbdelGawad2018b}---clearly
illustrates that interval types are more expressive (\emph{i.e.},
can express more types) than wildcard types.\footnote{Given that we do not present graphs of $S_{x3}$ or higher in the
remaining examples below (due to the very large size of these graphs),
this difference in expressiveness between interval types and wildcard
types is not as evident in later examples as it is in this example.}

It should also be noted that the subgraph highlighted with blue edges
in the graph of Figure~\ref{fig:IC2} (and some other unhighlighted
subgraphs of the same graph) seems to hint at how \emph{bounded} type
variables---with both upper and lower bounds---may be dealt with in
the full subtying relation, but we keep this research to future work.
\end{example}
\smallskip{}

\begin{figure}
\noindent \begin{centering}
\subfloat[$C_{2}$]{\noindent \protect\centering{}\protect\includegraphics[scale=0.5]{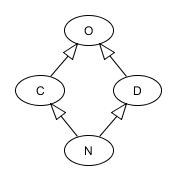}\protect}~~~~~~~~~\subfloat[$S_{21}=C_{2}\protect\pcgp_{\{\mathtt{C},\mathtt{D}\}}S_{20}^{\Updownarrow}$]{\noindent \protect\centering{}\protect\includegraphics[scale=0.5]{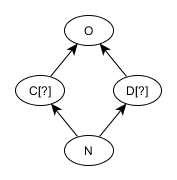}\protect}
\par\end{centering}

\subfloat[$S_{22}=C_{2}\protect\pcgp_{\{\mathtt{C},\mathtt{D}\}}S_{21}^{\Updownarrow}$]{\noindent \protect\centering{}\protect\includegraphics[scale=0.2]{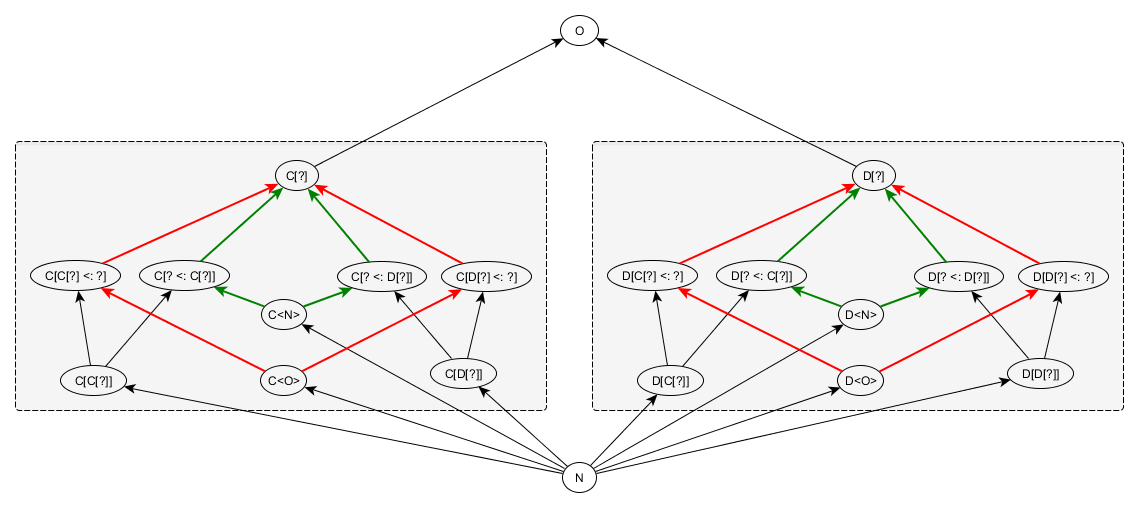}\protect}\protect\caption{\label{fig:Illustrating-the-use-2}Constructing generic OO subtyping
with interval types}
\end{figure}

\begin{example}
Consider the two Java class declarations

\code{\textbf{class} C<T>~\{\}}

\code{\textbf{class} D<T>~\{\}}.

\noindent The graphs in Figure~\ref{fig:Illustrating-the-use-2}
illustrate the construction of the subtyping relation $S_{2}$ corresponding
to these class declarations.
\end{example}
\smallskip{}

\begin{figure}
\noindent \begin{centering}
\subfloat[$C_{3}$]{\noindent \protect\centering{}\protect\includegraphics[scale=0.5]{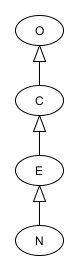}\protect}~~~~~~~~~\subfloat[$S_{31}=C_{3}\protect\pcgp_{\{\mathtt{C},\mathtt{E}\}}S_{30}^{\Updownarrow}$]{\noindent \protect\centering{}~~~~~~~~~~~\protect\includegraphics[scale=0.5]{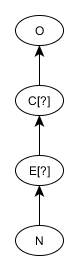}~~~~~~~~~~~\protect}
\par\end{centering}

\noindent \centering{}\subfloat[$S_{32}=C_{3}\protect\pcgp_{\{\mathtt{C},\mathtt{E}\}}S_{31}^{\Updownarrow}$]{\noindent \protect\centering{}\protect\includegraphics[scale=0.3]{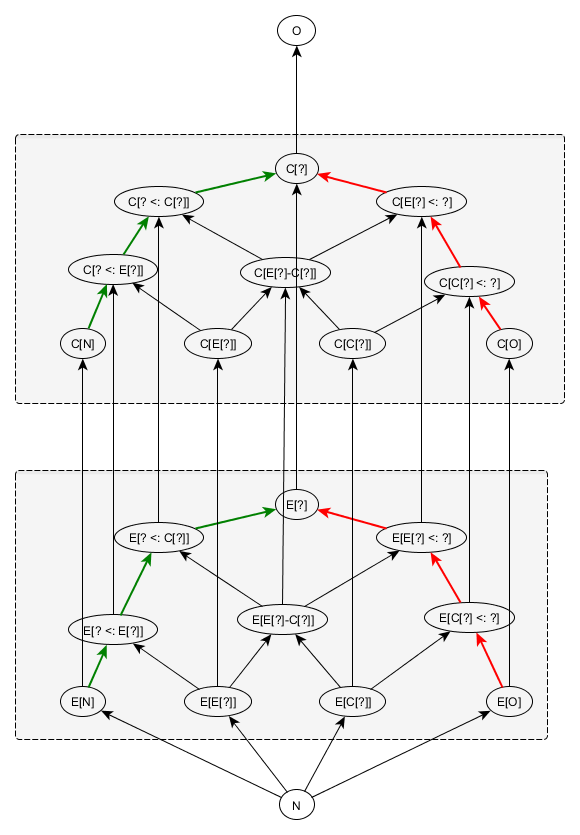}\protect}\protect\caption{\label{fig:Illustrating-the-use-3}Constructing generic OO subtyping
with interval types}
\end{figure}

\begin{example}
Consider the two Java class declarations

\code{\textbf{class} C<T>~\{\}}

\code{\textbf{class} E<T> \textbf{extends} C<T>~\{\}}.

\noindent The graphs in Figure~\ref{fig:Illustrating-the-use-3}
illustrate the construction of the subtyping relation $S_{3}$ corresponding
to these class declarations.
\end{example}
\smallskip{}

\begin{figure}
\noindent \begin{centering}
\subfloat[$C_{4}$]{\noindent \protect\centering{}\protect\includegraphics[scale=0.5]{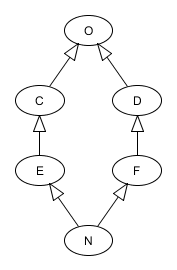}\protect}~~~~~~~~~\subfloat[$S_{41}=C_{4}\protect\pcgp_{\{\mathtt{F}\}}S_{40}^{\Updownarrow}$]{\noindent \protect\centering{}\protect\includegraphics[scale=0.5]{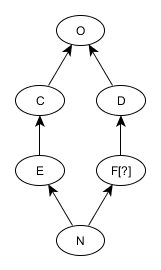}\protect}
\par\end{centering}

\noindent \centering{}\subfloat[$S_{42}=C_{4}\protect\pcgp_{\{\mathtt{F}\}}S_{41}^{\Updownarrow}$]{\noindent \protect\centering{}\protect\includegraphics[scale=0.3]{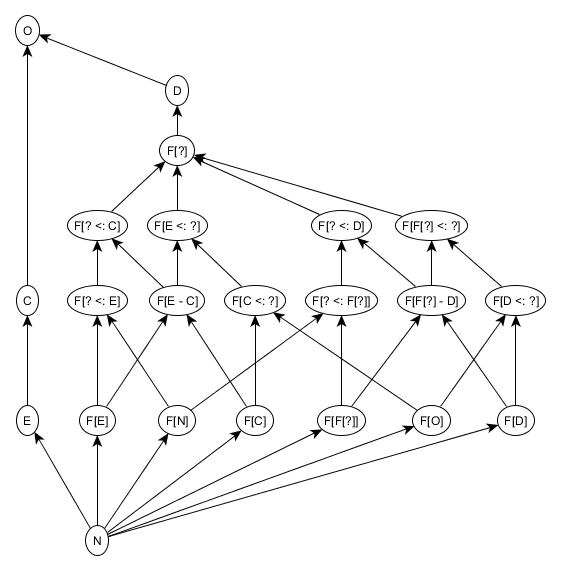}\protect}\protect\caption{\label{fig:Illustrating-the-use-4}Constructing generic OO subtyping
with interval types (automatic layout by yEd)}
\end{figure}

\begin{example}
Consider the four Java class declarations

\code{\textbf{class} C~\{\}}

\code{\textbf{class} E \textbf{extends} C~\{\}}

\code{\textbf{class} D~\{\}}

\code{\textbf{class} F<T> \textbf{extends} D~\{\}}.

\noindent The graphs in Figure~\ref{fig:Illustrating-the-use-4}
illustrate the construction of the subtyping relation $S_{4}$ corresponding
to these declarations. (As done in earlier examples via highlighting,
in this example the reader is invited to find out the subgraphs of
$S_{42}$ that are similar---\emph{i.e.}, isomorphic---to $S_{41}$
or its subgraphs, and to layout $S_{42}$ accordingly.)
\end{example}

In Appendix~\ref{sec:SageMath-Code} we present SageMath~\cite{Stein2017}
code that helped us produce some of the examples in this paper. We
used the code to produce larger examples of Java subtyping with interval
types, with larger numbers of generic class declarations and for higher
iteration indices than presented here. However, the very large size
of these graphs---with hundreds, thousands, and even tens of thousands
of vertices and edges---seems to diminish the usefulness of presenting
diagrams of these graphs.

However, a comparison of the sizes and orders of graphs corresponding
to Java subtyping wildcard types to those of graphs corresponding
to Java subtyping with interval types does reveal the greater degree
of expressiveness of interval types compared to the expressiveness
of wildcard types. The comparison of the sizes and orders of the subtyping
graphs also reveals that the difference in expressiveness is more
prominent when the subclassing relation (\emph{i.e.}, $C$, the basis
for defining the subtyping relation $S$ in both cases) has longer
paths (\emph{i.e.}, than those of the examples we present in this
paper. Much longer/deeper subclassing hierarchies are in fact typically
the case for most real-world OO programs) or for higher iteration
numbers in constructing the subtyping relation $S$ (\emph{i.e.},
for deeper-nested/higher-ranked generic types, which however are not
quite widely used in real-world OO programs).

\section{\label{sec:Related-Work}Related Work}

Although viewing wildcard types in Java as ``some sort of intervals''
seems to not be quite a new idea, but, other than presenting the idea
as a vague intuition that may help in understanding wildcard types,
it seems the idea has not been researched and presented more thoroughly
before.

As to other work that is related to the research we present here,
we already mentioned our concurrent work presented in~\cite{AbdelGawad2018b}
and our earlier work presented in~\cite{AbdelGawad2017a}, both of
which paved the way for the work presented here.

The addition of generics to Java has motivated much earlier research
on generic OOP and also on the type safety of Java and similar languages.
Much of this research was done before generics were added to Java.
For example, the work in~\cite{Bank96,Bracha98,Corky98} was mostly
focused on researching OO generics, while the work in~\cite{drossopoulou99,flatt99}
was focused on type safety.

Some research on generics was also done after generics were added
to Java, \emph{e.g.},~\cite{Zhang:2015:LFO:2737924.2738008,AbdelGawad2016c,Grigore2017,AbdelGawad2017b}.
However, Featherweight Java/Featherweight Generic Java (FJ/FGJ)~\cite{FJ/FGJ}
is probably the most prominent work done on the type safety of Java,
including generics. Variance annotations and wildcard types were not
put in consideration in the construction of the operational model
of generic nominally-typed OOP presented in~\cite{FJ/FGJ} however.

Separately, probably as the most complex feature of Java generics,
the addition of \textquotedblleft wildcards\textquotedblright{} (\emph{i.e.},
wildcard type arguments) to Java (in~\cite{Torgersen2004}, which
is based on the earlier research in~\cite{Igarashi02onvariance-based})
also generated some research that is particularly focused on modeling
wildcards and variance annotations~\cite{MadsTorgersen2005,KennedyDecNomVar07,Cameron2008,Summers2010,Tate2011,Tate2013,Greenman2014}.
This substantial work, particularly the latest work (of~\cite{Tate2011},~\cite{Tate2013}
and~\cite{Greenman2014}), clearly points to the need for more research
on wildcard types and generic OOP.

\section{\label{sec:Discussion}Discussion and Future Work}

In this paper we presented how type arguments for ground generic OO
types can be smoothly generalized from wildcard types to interval
types, thereby uniformly supporting types other than \code{Object}
and \code{Null} as upper and lower bounds for generic type arguments,
thereby subsuming and generalizing wildcard types. Also, our work
can be made more comprehensive\textemdash covering more features of
the Java generic subtyping relation\textemdash if it is extended to
model subtyping between generic types that have type variables in
them. We hinted earlier in this paper, in particular, to how bounded
type variables can be modeled, but including type variables and bounded
type variables in the construction of the Java subtyping relation
remains to be done.

We also believe it may be useful if a notion of \emph{nominal intervals}~\cite{AbdelGawad2016c}
(\emph{i.e.}, nominal type intervals) is supported in Java and other
similar generic nominally-typed OO programming languages, where type
variables (including bounded ones) can be viewed as names for interval
types (as presented in this paper) and where nominal intervals with
the same bounds but with different names are considered distinct (\emph{i.e.},
unequal) nominal intervals. We conjecture that the nominality of nominal
intervals can be a notion that is simpler to reason about than the
existentiality and ``capturing'' notions currently used in modeling
wildcard types. The fine details of this work, however, also have
yet to be decided and sorted out.

\bibliographystyle{plain}

\appendix

\section{\label{sec:SageMath-Code}SageMath Code}

In this appendix we present SageMath~\cite{Stein2017} code that
we used to help produce some of the graph examples presented in this
paper. The code presented here is not optimized for speed of execution
but rather for clarity and simplicity of implementation, and it builds
on and makes use of code presented in Appendix~A of~\cite{AbdelGawad2018b}.

\begin{lstlisting}[language=Python,basicstyle={\small\ttfamily},frame=lines]
IntvlFromTo = ' - '
ExtW = ExtStr+W

# is subpath/sublist/subinterval?
def is_sublist(sl, l):
  if sl == []:
     return True
  if l.count(sl[0]) == 0:
     return False
  start = l.index(sl[0])
  return (sl == l[start:start+len(sl)])

# interval string
def ty_intvl(path):
  f = path[0]
  l = path[len(path)-1]
  return W if f == BotCls and l == TopCls else
        (f if f == l else 
        (WExt+l if f == BotCls else 
        (f+ExtW if l == TopCls else
        (f+IntvlFromTo+l))))


# Use paths to get interval type arguments
# (subsumes Cov,Con,Inv)
def ITAs(S):
  ITA = DiGraph()
  ntps = S.all_simple_paths()
  # non-trivial paths, shortest to longest

  ntps.reverse() # from longest to shortest

  i=0
  for ntp in ntps:
    i=i+1
    lp = len(ntp)
    if lp > 2:
       shrtr_paths = S.all_simple_paths(
          max_length=lp-2, trivial=true)
       # most time-consuming step. constant
       # for all ntp of same length.

       # all_simple_paths doesn't work
       # properly if max_length == 0
       # (returns [] rather than trivial
       # paths only)
    else:
       shrtr_paths = map(lambda x: [x],
                         S.vertices())
    for sp in shrtr_paths:
      if is_sublist(sp, ntp):
         ITA.add_edge((ty_intvl(sp),
                       ty_intvl(ntp)))

  ITA = ITA.transitive_reduction()
  return ITA

ILP = '[' # LeftParen for intrvl type args
IRP = ']' # RightParen for intrvl type args

def ity(c,ita):
  return c+ILP+ita+IRP

def IntervalsSubtyping(subclassing, lngc,
                       FN_Prfx, num_iter):
  #Definition of SI0 (initial SI)
  SI0=subclassing.copy()
  SI0.relabel(lambda c: c if (c in lngc)
                        else ity(c,W))

  SI = SI0
  lst = [SI]

  for i in [1..num_iter]:
    ITA = ITAs(SI)

    # main step
    SI = GSP(subclassing, lngc, ITA, ity)
 
    lst.append(ITA)
    lst.append(SI)

  return lst
\end{lstlisting}

\end{document}